\documentclass[10pt,conference,compsocconf]{IEEEtran}

\usepackage{graphicx}
\usepackage{url}
\usepackage{balance}
\usepackage{xspace}

\usepackage[pdftex,colorlinks=true,pdfstartview=FitV,
 linkcolor=black,citecolor=black,urlcolor=black]{hyperref}

\begin{document}

\title{On Designing Better Tools for Learning APIs}

\author{\IEEEauthorblockN{Adrian Kuhn}
\IEEEauthorblockA{Software Practices Lab\\University of British Columbia}
\and
\IEEEauthorblockN{Robert DeLine}
\IEEEauthorblockA{Microsoft Research\\Redmond, WA}}

\newcommand{\ZoomableUML}{Zoomable UML}
\newcommand{\ConceptMap}{Concept Map}
\newcommand{\FacettedSearch}{Faceted Search}
\newcommand{\RichIntellisense}{Rich Intellisense}
\newcommand{\CloudREPL}{Interactive Snippets} 

\newcommand{\ie}{\emph{i.e. }}
\newcommand{\eg}{\emph{e.g. }}


\maketitle

\maketitle


\begin{abstract}\emph{
Modern software development requires a large investment in learning application programming interfaces (APIs).
Recent research found that the learning materials themselves are often inadequate: developers struggle to find answers beyond simple usage scenarios.
Solving these problems requires a large investment in tool and search engine development.
To understand where further investment would be most useful, we ran a study with 19 professional developers to understand what a solution might look like, free of technical constraints. In this paper, we report on design implications of tools for API learning, grounded in the reality of the professional developers themselves. 
The reoccurring themes in the participants' feedback were trustworthiness, confidentiality, information overload and the need for code examples as first-class documentation artifacts.
}\end{abstract}



\section{Introduction}

Modern software development requires a large investment in learning application programming interfaces (APIs), which allows developers to reuse existing components. API learning is a continuous process. Even when a developer makes a large initial investment in learning the API, 
the developer will continue to consume online material 
throughout the development process. These materials include reference documentation from the API provider, sample code, blog posts, and forum questions and answers. 

Indeed, seeking online API information has become such a pervasive part of modern programming that emerging research tools blend the experiences of the browser and the development environment. For example, Codetrail automatically links source code to the web pages viewed while writing the code \cite{gm09}. Blueprint allows a developer to launch a web query from the development environment and incorporate code examples from the resulting web pages \cite{bdwk10}. While these new tools help reduce the cost of (re)finding relevant pages and incorporating information from them, this covers only a portion of developers' frustrations. 
In a recent study of API learning obstacles among professional developers, Robillard found that the learning materials themselves are often inadequate~\cite{robillard09}. Bajracharya and Lopes analysed a year's worth of search queries and found that current code search engines address only a subset of developers needs~\cite{Bajracharya2009a}.   
Solving these systematic problems requires a large investment, either in the API provider's official documentation, the API users' community-based documentation, or in the search engines that unite the two \cite{bdwk10,Hoffmann2007a}. 
Any one of these changes is difficult and expensive.


To understand where further investment would be the most useful, we ran a study with 19 professional developers from Microsoft Corporation, with the goal of understanding what an ``ideal'' solution might look like, free from technical constraints. We invited randomly chosen members of a corporate email list of  Silverlight users to participate in one-hour sessions for small gratuities. Silverlight is a large API for creating web applications, with hundreds of classes for data persistance, data presentation, and multimedia. All participants were male with an average of $12.2$ years of professional experience. 

Borrowing from participatory design, we asked the participants to act as our partners in designing a new user experience for learning Silverlight. We ran two types of sessions.  In the first, we interviewed participants to investigate their common learning materials and most challenging learning tasks and then asked them to sketch a design for a new learning portal. We compiled these ideas into five exploratory designs. In the second type of session, we ran focus groups to get feedback on our descriptions of their learning tasks and the five designs. 

This paper's main contributions are a compilation of design implications for API learning tools, grounded in the reality of the professional developers themselves. We report on the recurring themes in the participants' feedback: trustworthiness, confidentiality, information overload and the need for code examples as first-class documentation artifacts.

\begin{figure*}
    \center
    \vspace{0.1in}
    \includegraphics[width=\linewidth,height=1.6in]{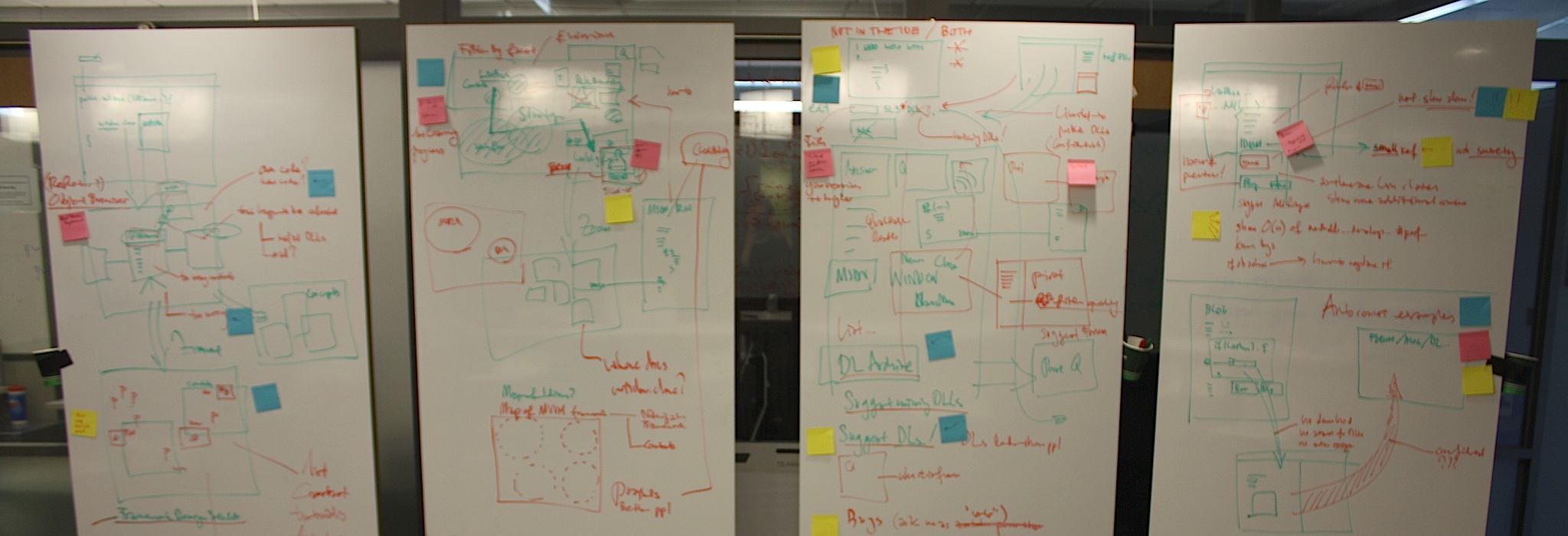}
    \caption{The five solution designs as narrated to the participants (from left to right): \ZoomableUML, \ConceptMap, \FacettedSearch, and on the last panel \RichIntellisense~(above) and \CloudREPL~(below). Pen color has been used to distinguish the designs (green) from the participant's input (red). The stick notes are the participant's votes.}
    \vspace{-0.2in}
    \label{fivedesigns}
\end{figure*}

\section{First Study: Current Practice}

In the first type of study session, we individually met with nine participants, and ran each through three activities. First, we asked the participant to describe all the  materials he used for learning Silverlight, as we recorded them on the whiteboard. Next, we asked him to consider this as a set of ``ingredients'' and to sketch a design for a learning portal that presents some or all of these ingredients to help developers learn Silverlight. Finally, we asked him to review the design by comparing the experience of learning Silverlight by using the design versus his own experience learning Silverlight.

\subsection{Learning Sources}
We asked the participant to describe all the  materials he used for learning Silverlight, as we recorded them on the whiteboard. Some of the learning sources are obvious and readily reported by participants, such as books and web search. To learn about non-obvious learning sources, we asked developers ``did you ever find an answer to a technological question that is not listed here,'' which led to answers like reverse engineering or social networking. Their reported learning sources are the following:

\textbf{``Off the top of my head''} seem to be the most common way developers find answers on the job. Most participants reported that they set aside dedicated time for learning. Typical off-the-job learning sources are: lurking on mailing lists and forums, watching videos and reading books. Most knowledge however is based on experience and acquired through learning-by-doing on the job. One participant refered to this a \emph{``growing your own folklore.''}

\textbf{Web search} was reported by all participants as the first place to go when they have an information need. Among the search results participants are typically looking for are: blog posts, discussion forums, official reference documentation, mailing list archives, bug reports and source repositories (listed in order of typical access patterns). 

\textbf{Intellisense} (\ie auto-completion of identifiers) was reported as a tool for the discovery of unknown APIs by all developers. One participant called this \emph{``digging through the namespaces.''} Discovering unknown APIs is an appropriation of auto-completion, originally conceived to help recall names from familiar APIs. 

\textbf{Prototyping,} reverse engineering and many more forms of tinkering were reported by all participants as a last resort when all above sources failed to provide an answer. 
Developers typically use prototyping both as an explorative tool and to verify hypotheses about the APIs. All participants reported that having to \emph{``get your hands dirty''} is an integral part of their learning experience. 

\textbf{Asking another person} was reported by most participants as a last resort. Developers follow a ``due diligence'' process before asking another person for an answer. It is important to them to have put enough personal effort into finding an answer before asking on a mailing list or reaching out to a person from their social network. 

These findings are consistent with Robillard's study of learning obstacles~\cite{robillard09}, but provide a more complete catalog of learning materials. Both studies found that developers strive to stay within the programming patterns and use cases that the API provider intends (even when that intent is undocumented) and that developers typically lack documentation when using an API for a less common task. Somewhat surprisingly, we found that developers prefer the community-based learning materials on the web, like blogs, forum posts, and tutorials, over more ``authoritative'' learning material, like books and reference documentation. Developers also prefer active but potentially time-consuming information seeking,~like iterative web search and reverse engineering, over possibly having to wait for answers by others, because they perceive the answers as more immediate.

\subsection{Learning Categories}


Based on the design sketches that participants produced, we elicited three broad categories of learning tasks:

\textbf{Technology selection} is learning about an API's fundamental capabilities (\emph{``Can Silverlight play video in this codec?''}) and comparing capabilities (\emph{``Is DirectX or Silverlight better for my game?''}). Sometimes the selection decision is about growing skills rather than project requirements. 

\textbf{Mapping task to code} includes both discovery of unfamiliar APIs as well as remembering relevant identifier names in previously learned APIs. Getting an answer to this type of questions typically falls in two phases. Initially developers search based on natural language task descriptions (e.g. \emph{``how to implement a zoomable canvas''}) and skim through many search results to stumble on relevant identifier names. Once they have a concrete identifier, their search behavior becomes more focused and may be as simple as looking up the identifier's reference documentation.

\textbf{Going from code to better code} is a major concern of professional developers. All participants reported that they spend considerable effort getting answers to performance questions. Other use cases are robustness and idiomatic, \ie intended use of an API, in particular, breaking changes of newly released API versions or different target platforms.


\section{Second Study: Solution Elements}

For the second type of session, we compiled the user feedback from the first sessions into five exploratory designs. We ran three focus groups of totally 10 participants  (with three, three, and four members) and asked them to provide feedback on the five designs. In each session, we drew each design on its own whiteboard and encouraged participants to ask questions, provide feedback, and to add their own ideas as we explained the design. 

Figure~\ref{fivedesigns} shows a photograph of the whiteboards with the five designs, taken at the end of a focus group's session. In the following, the designs are described in the order they were presented to the participants in that session:

\paragraph{Design: \ZoomableUML} 
This design draws from the spatial software representation of CodeCanvas~\cite{Deline2010a} and addresses answering complex reachability questions~\cite{Latoza2010a} as you code. The design extends the IDE with a zoomable UML diagram. The diagram is zoomed on locally reachable types of the API and shows their dependencies and interaction. The user can zoom out to get a larger picture of the API, up to the level of namespaces.

\paragraph{Design: \ConceptMap} 
The API is presented as a zoomable map, organized around programming domain concepts (e.g. ``controls'', ``media content''). As the user zooms in, the concepts become more refined (e.g. ``streaming video''). At the lowest zoom level, the map shows web-based content about that concept, including blogs, forum posts, tutorials, and the people who author these.
The map is searchable and keeps track of user interaction as well as the user's learning progress. Users can bookmark locations and share their bookmarks. Documentation editors can use the same feature to share tutorials as ``sight-seeing tours.''

\paragraph{Design: \FacettedSearch}
This design unifies web search and asking people questions. The user types a question into a textbox. As she types, related  search results are pulled in from various sources (web sites, bug reports, code examples, mailing list archives, etc). Search results are grouped by facets, such as type of sources, type of content or semantic concepts. Besides the results, a tag cloud appears with extracted identifier names. Search results are summarized using code examples, if possible. In addition, the results include suggested people and mailing lists, to help developers reaching out to experts on the topic of the questions.

\paragraph{Design: \RichIntellisense} 
This design extends auto-completion of identifiers with search results that are automatically pulled from the world wide web. The results are ``localized'' to the current context of the IDE, such as imported libraries and reachable types \cite{Holmes2005}. Results are shown in the same pop-up windows as the auto-completion suggestions. If possible search results are displayed as code examples, ready to be incorporated into the code, as in Brandt et al \cite{bdwk10}.

\paragraph{Design: \CloudREPL} 
This design attaches an execution context to code examples on the web. Code examples include hidden meta-information with all context that is required to execute. Examples are editable, debuggable and can be executed live in the browser. With a single click, users can download examples into their IDEs. Similarly, users can upload the code in their IDE as runnable examples on the web, for inclusion in blogs or discussion forums.

\section{Feedback}

After we explained all five designs, we then handed each participant a pen and sticky notes and gave them 10 minutes to annotate the designs, either with a blank sticky note to mean ``I like this part'' or with their own comments (typical ones were smiley faces, frowny faces, ``NO'', etc). The most popular designs are ``\FacettedSearch`` and for learning activities the ``\ConceptMap`` design. Participants downvoted the ``\ZoomableUML'' and ``\RichIntellisense'' due to concerns about information overload, the same happened with ``\CloudREPL`` due to concerns about missing confidentiality.

There were several recurring themes in our participants' feedback which cut across the various designs. The four top most recurring themes are discussed and summarized as design implications for tool builders in the following:

\subsection{Code Examples}

We got very positive feedback on the emphasis on code examples and identifier names in the ``\FacettedSearch'' design. Participants prefer results with code examples over results without code examples, which is supported by existing research on API learning barriers \cite{robillard09}. When mapping a task to code, developers typically use web search and linearly go through all results until they find one with a code example or an identifier; often repeating this process a dozen times until they find a working answer. Participants liked the facetted search design for the extracts code examples and identifiers from top search results. One participant even said that the summary tag cloud with identifiers, by itself, would be reason to use it.

\emph{Implication for tool builders:} Developers need the heterogeneous learning materials that web search provides, but want it to be more targeted and organized. Search engines for API learning should extract code examples and identifiers found in natural text documents, and present them to the developers in a more accessible way. This implication is supported related work on code examples \cite{bdwk10,Hoffmann2007a,Holmes2005}.

\subsection{Credibility}

Credibility of web sources appeared as a major concern with all designs that included content taken from the web. For the participants, credibility is mostly a function of where the information comes from. For example, participants reported that search results from blogs are often more relevant, but typically less credible than official reference documentation. They also rely on the social reputation of its source rather than technical factors, which supports existing research \cite{Gysin2010a}. In particular with the ``\FacettedSearch'' design, which automatically summarizes search results, participants emphasized the importance of seeing the information source to judge credibility.  

\emph{Implication for tool builders:} Tools should show both credibility and relevance when presenting search results, such that the developers can make an informed decision when using API information and code examples from the web. To asses the credibility of API information tools should prefer social factors, such as the credibility of the information's author, rather than technical statistics, such as code metrics.

\subsection{Confidentiality}

Confidentiality appeared as a major concern with all designs that share local information with a global audience. In particular with the ``\CloudREPL'' design, which publishes an example's execution context on the web, participants were concerned with leaking proprietary information, like the use of certain libraries. One participant was also concerned that publicly inquiring about technologies could accidentally reveal business strategies.

\emph{Implication for tool builders:} When automatically sharing local information with the web, tools must be careful about protecting proprietary information, such as not showing confidential code, nor libraries being used. Tools should give developers full control over shared information, for example by letting them review the list of automatically included terms before issuing the search query. Or alternatively, only sharing information that is on a user controlled white list.

\subsection{Information Overload} 

Information overload was the major reason why participants rejected the ``\ZoomableUML'' and the ``\RichIntellisense'' designs. We got strong feedback that pulling more information into the IDE is not welcome unless it is highly task- and context specific information. Participants were also concerned that adding more features to Intellisense's popup will use too much screen real estate and slow down the IDE. 

\emph{Implication for tool builders:} Any tool that pulls additional information into the IDE must be highly selective and should only show information that is specific to the developer's current task and context. The ability to further filter down the information is crucial, as well as not slowing down the IDE and using screen real estate sparingly.





\subsection{Threats to validity}

We selected all participants from the same corporation, whose common hiring practices and corporate culture may bias the results. In particular, the participants all work for the same company that produces the Silverlight API, which gives the participants unique access to the API creators. 
Nonetheless, participants mostly accessed public learning outside the company and many expressed hesitation about asking questions of fellow employees for fear of harming their reputation. 
The study is also based on a single API. While this choice allowed us to compare participants' experiences and gave them common ground during the focus groups, there may be issues in learning Silverlight that do not generalize to other APIs.

\section{Conclusion}



Web search is the predominant form of information seeking, but in many cases is frustrating and error-prone. Developers need the heterogeneous learning materials that web search provides, but want it to be more targeted and organized. 
Therefore, API learning tools that bring web search and development environments closer together should
	1)~leverage examples and identifiers found in natural text documents as first-class results,
	2)~communicate the credibility of aggregated results,
	3)~filter search results by task and context to avoid information overload,
	but should avoid 3)~sharing confidential information without the user's consent.


\small{Acknowledgments: The first author is grateful to MSR for the unique experience of being a research intern in 2010. We thank C. Albert Thompson for proof-reading a final draft of this paper.}

{\small
\bibliographystyle{abbrv}
\bibliography{api-learning}}

\end{document}